\documentclass[11pt,nofootinbib,prd]{revtex4}
\usepackage[latin9]{inputenc}
\usepackage{graphicx}
\usepackage{epstopdf}
\usepackage{amsmath}
\usepackage{color}
\usepackage{subfigure}

\makeatletter
\usepackage{slashed}
\makeatother

\def\lb#1{\if 1#1 \ln\beta \else \ln^#1\beta \fi}
\def\lt#1{\if 1#1 \ln 2 \else \ln^#1 2 \fi}

\newcommand{\be}{\begin{equation}}
\newcommand{\ee}{\end{equation}}
\newcommand{\ba}{\begin{eqnarray}}
\newcommand{\ea}{\end{eqnarray}}

\newcommand{\ep}{\epsilon}

\newcommand{\ttbj}{t\bar{t}j}

\begin{document}

\title{
Top quark pair production in association with a jet: \\QCD corrections 
and  jet radiation  in top quark decays
}

\author{Kirill Melnikov$^1$, Andreas Scharf$^{2,3}$ and Markus Schulze$^{1,4}$ }

\affiliation{$^1$
Department of Physics and Astronomy, Johns Hopkins University, Baltimore, MD, USA
}
\affiliation{$^2$
Department of Physics, State University of New York at Buffalo,
Buffalo, NY, USA
}
\affiliation{$^3$
Institute for  Theoretical Physics and Astrophysics, 
University of W\"urzburg, W\"urzburg, Germany
}
\affiliation{$^4$
High Energy Physics Division, Argonne National Laboratory, Argonne, IL 60439, USA
}

\makebox[6.5in][r]{\hfill ANL-HEP-PR-11-76}\\

\begin{abstract}
We consider top quark pair production in association with a hard jet through next-to-leading order in perturbative QCD.
Top quark decays are treated in the narrow width approximation and spin correlations are retained throughout the computation.
We include hard jet radiation by top quark decay products and explore their importance for basic kinematic distributions at the Tevatron and the LHC.
Our results suggest that QCD corrections and jet radiation in decays can lead to significant changes in shapes of basic distributions and, therefore,
need to be included for the description of $t \bar t j$ production.
We compare the shape of the transverse momentum distribution of a top quark pair recently measured by the D0 collaboration with the
result of our computation and find reasonable agreement. 
\end{abstract}
\maketitle

\section{Introduction}

Experiments at the  LHC are in the process of accumulating 
a large data set of top quark pairs  that will allow detailed 
studies of various processes that Tevatron experiments either 
observed  with relatively low statistics  or did not observe at all.   
Such processes  include associated 
production of a $t \bar t$ pair with a jet \cite{joey}, 
a photon \cite{cdf},  two jets, a $Z$-boson
 or a Higgs 
boson.   Beyond studies of $t \bar t$ pair production at very high 
invariant masses,  detailed investigations  of associated production  processes 
will mark the beginning of the post-Tevatron era in top quark physics. 
A significant body of theoretical work is devoted to improving 
predictions for $t \bar t$ associated production 
processes, see  Refs.~\cite{Lazopoulos:2008de,Kardos:2011na,Garzelli:2011is,
Frederix:2011zi, Dawson:2002tg,Beenakker:2002nc,Melnikov:2011ta,Bevilacqua:2010ve,Bredenstein:2010rs,Bevilacqua:2009zn}.

It is well-known that, once produced, top quarks decay very rapidly.
For this reason  top quarks 
are observed and studied indirectly through kinematic features
of their decay products.
Unfortunately, this complicates
top quark studies by introducing 
additional uncertainties in kinematic
reconstructions due to finite resolution on energies and angles
of decay products, missing energy as well as  backgrounds,
including combinatorial ones.  
On the positive side, the rapid  decay of top quarks
enables the description of their decay products in perturbative QCD without
the need to resort to fragmentation functions and other non-perturbative 
objects. 

A precise
 description
of hard hadron collisions requires the application of
 perturbative QCD through 
next-to-leading order (NLO) in the expansion of the strong coupling
constant. The complete 
NLO QCD description of any process that 
involves $t \bar t$ production should include 
QCD corrections to top  quark pair 
production {\it and} to top quark decays.
For processes where top quarks are produced in association with
a photon or a jet,  
a standard  process to study is $t \bar t X$ production 
with $X = \gamma,j$, followed by the top quark decay $t \to b W$. However, 
since both photons and jets can be radiated in top quark decays, 
one should also consider $t \bar t$ production followed by 
``radiative'' decays, such as  $t \to bWj$ and $t \to bW \gamma$.  
The importance of radiation in the decays strongly depends on the selection 
criteria that are used to isolate a particular process and, hence, can not 
be quantified a priori.  For example, in a recent
measurement of $t \bar t \gamma$ production
by the CDF collaboration \cite{cdf}, about {\it half} 
of all signal events come
from the process $p\bar p \to t \bar t$ followed by the radiative decay of the 
top quark $ t \to W b \gamma$ \cite{Melnikov:2011ta}.
To compare their measurement with theoretical predictions,
CDF uses a NLO QCD $K$-factor for the process $p \bar p \to t \bar t \gamma$
computed with {\it stable top quarks}. However, since about half of their
events come from $t \bar t$ production followed by radiative decays
of top quarks,  it is unclear if such a  comparison 
is meaningful.

In principle, one can get around the problem of
separating
production and decay stage 
by  simply giving up on the approximation
that 
{\it top quarks are produced on-shell} and focusing instead on the 
fully realistic
final state such as $b \bar b W^+W^- X$ with 
$X = \gamma, j, jj, H, Z$.  A calculation of  $pp \to b \bar b W^+W^- X$ 
through a given order in  the perturbative expansion in QCD
leads to a prediction for a final state that includes both
``resonant'' and ``non-resonant'' contributions, providing a complete
description of the process.  Without a doubt, this is the 
best approach possible, provided that it is feasible. 
The feasibility depends on the  approximation in perturbative 
QCD at which the process of interest is 
considered. At leading order, this approach can be pursued for essentially
arbitrarily complicated process thanks to automated programs such 
as Madgraph \cite{Stelzer:1994ta}. 
However, this approach becomes very complex already at NLO QCD.
For the {\it simplest} process
$pp \to W^+W^-b \bar b$ that, among many other ways, can occur through 
the production of a nearly on-shell $t \bar t$ pair, this was recently
accomplished  in Refs.~\cite{Denner:2010jp,Bevilacqua:2010qb}.
Applications of
this approach to more complicated processes are 
difficult  to imagine.    On the contrary, a sequential treatment 
of various production and decay stages 
based on the  double resonant approximation  for $t$ and $\bar t$
can be generalized to processes of significant complexity,
at least as a matter of principle. This double
resonance approximation is parametrically controlled 
by the ratio of the top quark width to its mass $\Gamma_t/m_t \sim 10^{-2}$
and should be sufficiently accurate for most observables.
In fact, there has been significant
progress in using this approximation to describe top quark pair production
recently.  
For example, 
$t \bar t$ pair production at NLO QCD in 
the double resonance  approximation, including corrections to top quark decays
and spin correlations, 
was computed in 
Refs.~\cite{Bernreuther:2001bx,Bernreuther:2001rq,Bernreuther:2001jb,brand,
Bernreuther:2004jv,Bernreuther:2004wz,ms,Bernreuther:2010ny}. 
The number of
similar computations for   more complicated processes is rather limited.
The only process for which a full  
description is available is associated production
of $t \bar t \gamma$   \cite{Melnikov:2011ta}, where NLO QCD corrections 
to the production and decays, including the radiative one ($t \to Wb\gamma$), are computed.

 The  production of $t \bar t j$ at NLO QCD was first studied in 
Ref.~\cite{Dittmaier:2007wz,Dittmaier:2008uj}
for stable top quarks and later in Ref.~\cite{Melnikov:2010iu} where decays were included at leading order.  
A different approach to  this process is  described in 
Refs.~\cite{Kardos:2011qa,alioli}, where $t\bar tj$ production at NLO QCD
is combined with a parton shower, following the POWHEG procedure
\cite{Nason:2004rx}. Top quark decays are treated in the parton 
shower approximation where 
$t \bar t$ spin correlations are omitted either  at leading 
\cite{Kardos:2011qa} or at next-to-leading \cite{alioli} order,
and whose correspondence with NLO QCD computations
is not clear.   

Fortunately, these approximations are not necessary, since 
it is possible to treat the complete process $t \bar t j \rightarrow b \bar{b} W^+ W^- j$ in the narrow width approximation
where top quark decays, including $t \to Wb j$, are described consistently 
at  NLO QCD and spin correlations are retained throughout the entire decay chain.
Such a calculation gives a state-of-the-art
description of the $t \bar t j$ production that, in principle, can be 
directly compared to experimental results because
theoretical predictions 
for a complete and fully realistic final state become  available.
The goal of the present paper is therefore 
to extend the description  of $pp \to t \bar tj$ production given 
in Ref.~\cite{Melnikov:2010iu} by including radiation in the decay 
through next-to-leading order  in perturbative QCD.

The paper is organized as follows. In the next Section, we
outline the framework  of our calculation and  
discuss technical 
aspects of the computation which arise because of the 
need to treat radiative 
corrections to processes with {\it decay} kinematics.
Phenomenological results for the Tevatron and the 7~TeV LHC are
presented in Section 3.
We conclude in Section 4.

\section{Technical aspects of the calculation}

In this Section, we summarize the technical aspects of the calculation.
We begin by describing  various
contributions that we require for the computation. As we pointed out 
already, the top quark is treated in the narrow width approximation. 
This allows us to organize the computation in 
terms of a production process which includes the hard collision,
and the decay process.
\\
To give a complete list of all necessary contributions 
for $t\bar{t}+{\rm jet}$ production calculation, we begin by writing the
formula for the inclusive cross-section
as a convolution of the 
production cross-section $\sigma_{t\bar t}$ and the decay rate $\Gamma_t$
\be
{\rm d} \sigma_{\rm incl} = \Gamma_{t,\rm tot}^{-2}
\left ( {{\rm d} \sigma_{t\bar t+0j} + {\rm d}\sigma_{t\bar t+1j}+ {\rm d}\sigma_{t\bar t+2j}+...} \right )
\otimes \left (
{\rm d} \Gamma_{t\bar t+0j} + {\rm d} \Gamma_{t\bar t+1j} + {\rm d}\Gamma_{t\bar t+2j}+...\right ).
\label{eq:inclu}
\ee
Subscripts denote the number of {\it exclusive} jets defined according to some jet algorithm.
We further use the abbreviation 
${\rm d}\Gamma_{t\bar t+nj} = \sum_{l=0}^{n} {\rm d}\Gamma_{t+lj} \; {\rm d}\Gamma_{\bar t+(n-l)j}$
to summarize the decay rates of top and anti-top quark in association with a 
fixed number of jets.

We can now expand Eq.(\ref{eq:inclu}) assuming that the number of jets that we eventually
require is equal or larger than one and that the
cross-sections and widths for each jet multiplicity scale
as $\sigma_{t\bar t,nj} \sim {\cal O}(\alpha_s^{2+n})$ and $\Gamma_{t,nj}
\sim {\cal O}(\alpha_s^n)$.
Since we are interested in NLO QCD corrections to one-jet production,
we can disregard all terms that  depend on powers of $\alpha_s$ higher than four.
We obtain
\be
{\rm d} \sigma_{t\bar t+1j}^{\rm NLO} = \Gamma_{t,\rm tot}^{-2} \Big(
{\rm d}\sigma_{t\bar t+0j} {\rm d}\Gamma_{t\bar t+1j}  + {\rm d}\sigma_{t\bar t+0j} {\rm d}\Gamma_{t\bar t+2j}
+ {\rm d}\sigma_{t\bar t+1j} {\rm d}\Gamma_{t\bar t+0j}
+ {\rm d}\sigma_{t\bar t+1j} {\rm d}\Gamma_{t\bar t+1j}
+ {\rm d}\sigma_{t\bar t+2j} {\rm d}\Gamma_{t\bar t+0j} \Big),
\ee
\begin{figure}[t]
  \subfigure[~jet emission in production] {\label{FigProd}
          \includegraphics[scale=0.4]{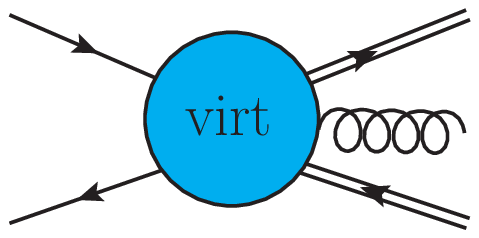} \hspace{5mm}
          \includegraphics[scale=0.4]{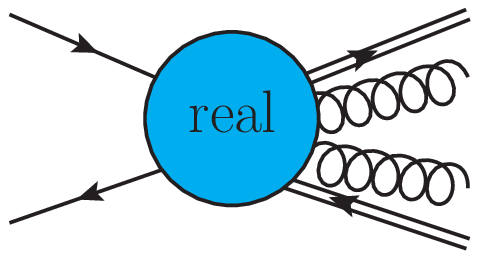} }
  \hspace{2cm}            
  \subfigure[~jet emission in decay]{\label{FigDec} 
          \includegraphics[scale=0.55]{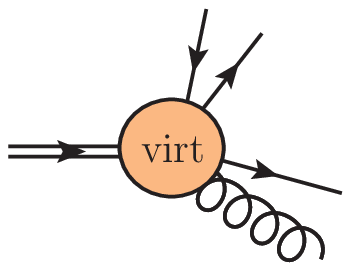} \hspace{5mm}
          \includegraphics[scale=0.55]{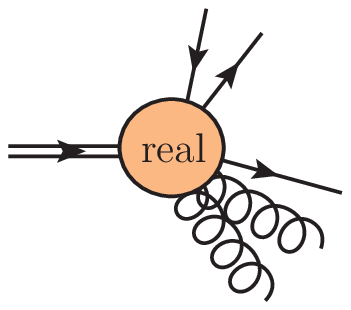}  }
  \\
  \subfigure[~mixed contribution]{\label{FigMix} 
          \includegraphics[scale=0.45]{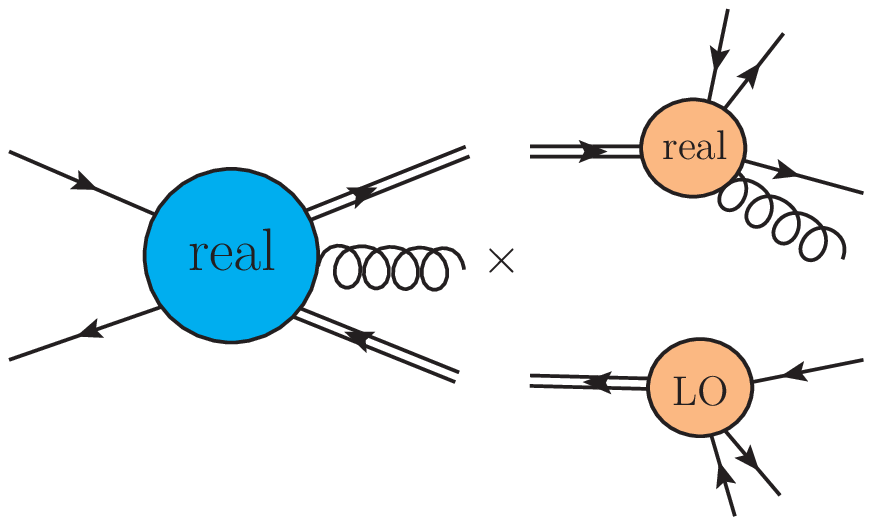} \hspace{1cm}
          \includegraphics[scale=0.45]{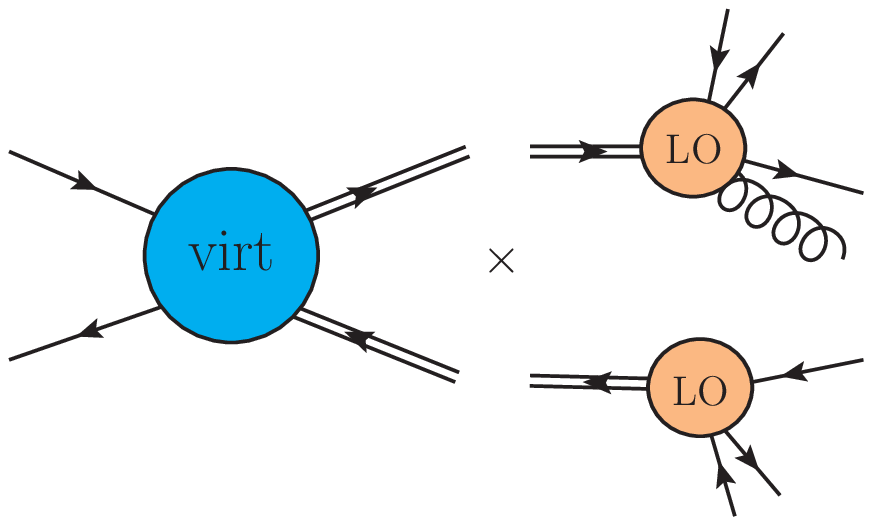}  \hspace{1cm}
          \includegraphics[scale=0.45]{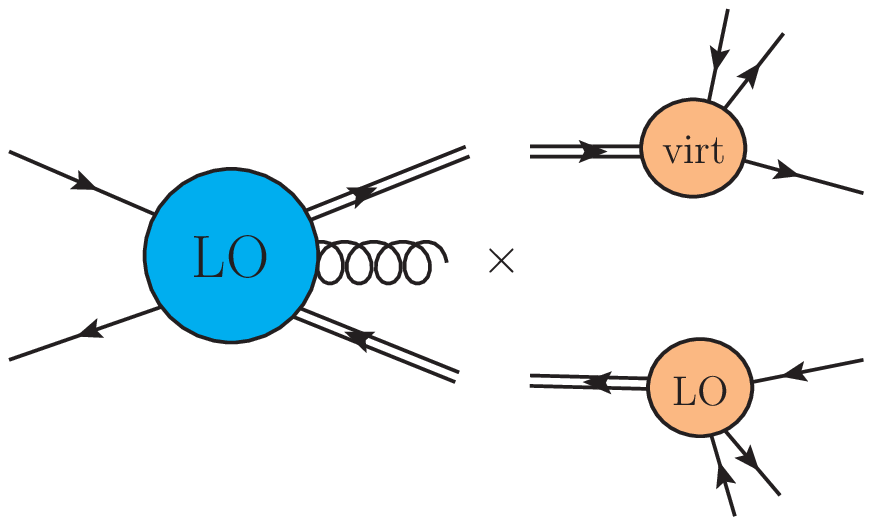}  }
\\
\caption{NLO QCD corrections to  top quark pair 
production and decay in association with a jet. 
Contributions (a) and (b) show jet emission in 
production and decay, respectively. The symbol ``real'' 
indicates that one parton is allowed to be unresolved.
(c) defines the ``mixed''~contributions.}
  \label{Fig21}
\end{figure}
and we re-write this formula in a way that separates various
processes that contribute to the cross-section \\[-1cm]
\begin{align}
{\rm d} \sigma_{t\bar t+1j}^{\rm NLO} =
\Gamma_{t,\rm tot}^{-2} & \Big( 
    {\rm d}\sigma^{\rm LO}_{t\bar t+1j} {\rm d}\Gamma^{\rm LO}_{t\bar t}
   + {\rm d}\sigma^{\rm LO}_{t\bar t} {\rm d}\Gamma^{\rm LO}_{t\bar t+1j}
   + \overbrace{ \big( {\rm d}\sigma^{\rm virt}_{t\bar t+1j} + {\rm d}\sigma^{\rm real}_{t\bar t+2j} \big){\rm d}\Gamma^{\rm LO}_{t\bar t} }^{{(a)}}
 \label{eq20} \\  \nonumber &
   + \underbrace{ {\rm d}\sigma^{\rm LO}_{t\bar t} \big( {\rm d}\Gamma^{\rm virt}_{t\bar t+1j} + {\rm d}\Gamma^{\rm real}_{t\bar t+2j}  \big) }_{{(b)}}
   + \underbrace{  {\rm d}\sigma^{\rm real}_{t\bar t+1j} {\rm d}\Gamma^{\rm real}_{t\bar t+1j}
   +  {\rm d}\sigma^{\rm virt}_{t\bar t} {\rm d}\Gamma^{\rm LO}_{t\bar t+1j} 
   + {\rm d}\sigma^{\rm LO}_{t\bar t+1j} {\rm d}\Gamma^{\rm virt}_{t\bar t}
     }_{{(c)}}
   \Big).
\end{align} 
We now review different contributions that appear in Eq.(\ref{eq20}).
The first and second term describe $t\bar t+j$ production 
at leading order followed by leading order decays of the top quark 
and  $t \bar t$ production  followed by a radiative decay  
of the top quark, respectively.
The third term represents the NLO QCD correction to 
the production process $t\bar t+j$, where the symbol ``real'' 
indicates that one parton is allowed to become unresolved.
The first term in the second line of Eq.(\ref{eq20}) describes leading order production of a top quark pair 
followed by NLO QCD corrections to the ``radiative decay'' $t \to W+b+j$
\footnote{We note that, in the case of a semi-leptonic 
top quark decay, also the W-boson is allowed to radiate an 
additional hard jet at NLO QCD. We include this contribution in our computation as well.}.
Finally, the last three terms describe mixed contributions 
where jet emission occurs simultaneously
in both production and decay stage. Since one of those jets can be unresolved, 
the last two terms are the corresponding virtual 
corrections needed to provide an infra-red finite  result.
In the remainder of the paper we will refer to contribution 
$(a)$ and $(b)$ in Eq.(\ref{eq20}) as {\it jet radiation
in the production} and {\it jet radiation in the decay}, 
respectively. The last part $(c)$ we call the {\it mixed} contribution.
The corresponding topologies are depicted in Fig.~\ref{Fig21}.

Let us now describe how NLO QCD corrections to jet radiation in the production processes  
$pp \to t \bar t $ and $pp \to t \bar t j$ are treated. We note that
-- when production processes are considered at next-to-leading order -- 
the decay processes are included at leading order, 
consistent with the expansion in $\alpha_s$.
However, these leading order decays are different processes: in 
the former case, we consider the radiative decay $t \to W b g$, 
since an additional jet is required in the final state. In the latter case, 
top quarks decay into the $Wb$ final state since the 
jet is created in the production stage. 
The  NLO QCD results for the production processes are available; they are 
described in Refs.~\cite{ms,Melnikov:2010iu}
including an efficient way of implementing the decays of top 
quarks while retaining all spin correlations.
We note  that 
one-loop QCD corrections to  $ 0  \to q \bar q  t\bar t$,
$0 \to gg t \bar t$, $0 \to q \bar q t \bar t g$ and 
$0 \to ggg t \bar t$  amplitudes  that we require   are calculated 
using generalized $D$-dimensional unitarity \cite{Giele:2008ve,egkm,
Ellis:2011cr}. 
The real emission corrections are obtained following the Catani-Seymour 
dipole subtraction formalism \cite{Catani:1996vz} and 
its extension to massive particles in Ref.~\cite{Catani:2002hc}.
To improve the efficiency of the computation, we 
follow Ref.~\cite{nagy} and
use $\alpha$-parameters to restrict subtraction terms to singular phase-space regions. 
The relevant dipoles with $\alpha$-parameters are found 
in Refs.\cite{Bevilacqua:2009zn,mcfm,Campbell:2005bb}.

The second part required in the calculation involves  leading order 
production processes $pp \to t \bar t$ and $pp \to t \bar t j$ 
followed by top quark decays at next-to-leading order. In the former 
case the NLO QCD corrections to radiative decays $t \to Wbj$ and $W\to q \bar q g$ 
are required; in the latter case $t \to Wb$ and $W\to q \bar q$ 
need to be computed through NLO
QCD.  
Radiative corrections to $t \to Wb$ and $W\to q \bar q$ are known; 
our implementation follows the description 
in Ref.~\cite{ms}. 
We do not repeat it here and focus, instead, on the 
NLO QCD corrections to the ``radiative decay'' $t \to bWj$. 
Since this is a sufficiently low-multiplicity process, 
we compute the virtual corrections 
using  Passarino-Veltman reduction of tensor integrals
\cite{pv}.  The scalar  integrals are taken from Ref.~\cite{Ellis:2007qk}.
For the calculation of the real corrections we need to consider various 
decay processes, such as 
 $t\!\to\! (W\!\to\! q\bar{q}')\, b gg$, $t\!\to\! (W\!\to\! q\bar{q}' gg) b$ and 
$t\!\to\! (W\!\to\! q\bar{q}' g) b g$ etc. 
The real emission subtraction terms are again constructed 
using the dipole formalism of Catani and Seymour~\cite{Catani:1996vz}. 
However, 
we note  that its application to decay processes 
requires clarification.  Catani and Seymour constructed subtraction terms -- 
the dipoles -- that satisfy two criteria: 1) they remove   infra-red 
and collinear singularities when subtracted from  scattering amplitudes 
and 2) they can be integrated analytically over the unresolved phase-space. 
In the original paper~\cite{Catani:1996vz},
it is shown how to satisfy these conditions for 
two colliding massless partons.
Since decay 
kinematics differ from production kinematics, 
some of the Catani-Seymour dipoles need to be modified
if we deal with decays of color-charged  particles.

Recall that within the Catani-Seymour dipole formalism, dipoles are 
constructed by taking  different partons 
to be ``emittors'' and  ``spectators'', in addition 
to soft or collinear  partons  that are actually  ``emitted''.  
The dipoles depend on 
``flavors'' (quarks, gluons) of ``emitted'' and ``emittors'' and 
on   whether ``emittors'' and ``spectators'' 
are in the initial or in the final state.
The corresponding dipoles are referred to as final-final,
final-initial, initial-initial and initial-final.

However, only a limited number of these dipoles 
is needed for the decay processes in general.
First, it is obvious that there are no initial-initial dipoles since 
there is just one particle in the initial state. Final-final dipoles 
can be borrowed from Ref.~\cite{Catani:1996vz} and the phase-space re-mapping therein.
Initial-final dipoles can be
omitted since real radiation by a massive initial state particle
is only singular in soft kinematics. This contribution can be 
absorbed into final-initial dipoles which 
are the only dipoles for decay kinematics that need to be constructed.

The complete list of dipoles that we need for the process
$t \to W\, b\, g_1\, g_2$ are 
$\mathcal{D}_{g_1 g_2, b}, \, \mathcal{D}_{b g_1, g_2}, \, \mathcal{D}_{b g_2, g_1},
 \mathcal{D}_{b g_1}^t, \,\mathcal{D}_{b g_2}^t$ and  $\mathcal{D}_{g_1 g_2}^t$.
The first three dipoles are of the final-final type whereas 
the last three dipoles  are the missing final-initial dipoles. 
We will discuss their construction in the following.
We need to distinguish two types of final-initial 
dipoles which correspond to the splitting
$q\to q g$ and $g \to g g$ with a top quark in the 
initial state being  the spectator.

We begin our discussion with the gluon-quark dipole.
It can be  extracted from Ref.~\cite{Campbell:2004ch}.
To this end, we consider the process $t \to W b g_1 g_2$ and imagine 
that gluon $g_1$ and the (massless) $b$-quark become unresolved. 
The top quark in the initial state is the spectator. We combine 
the momenta of the $W$-boson and  the gluon  $g_2$ into a new momentum 
${\tilde p}_{W} = p_W + p_{g_2}$ and introduce a variable 
$r^2 = {\tilde p}_W^2/m_t^2$.  The remaining momenta  -- whose scalar 
products lead to soft and collinear singularities -- are parametrized 
using two variables $z$ and $y$
\be
p_b p_{g_1} = \frac{m_t^2}{2} (1 - r)^2 y,\;\;\;\;\;
p_t p_{g_1} = \frac{m_t^2}{2} ( 1- r^2)(1-z).
\label{eq4}
\ee
With this  parametrization, the  final-initial gluon-quark 
dipole reads  \cite{Campbell:2004ch} 
\be
D_{g_1 b}^{t} = 4 \pi \alpha_s \mu^{2\epsilon} 
\left [ \frac{1}{p_b p_{g_1}} \left ( \frac{2}{1-z} - 1-z
- y \epsilon (1-z) \right ) - \frac{m_t^2}{(p_t p_{g_1})^2}
\right ] \delta_{\lambda \lambda'},
\label{eq5}
\ee
where $\ep = (4-d)/2$ is the parameter of dimensional
regularization, $d$ is the number of space-time dimensions 
and $\lambda,\lambda'$ are quark helicity labels. 
We note that Eq.(\ref{eq5}) gives the dipole in conventional 
dimensional regularization (CDR) scheme; if four-dimensional helicity (FDH) 
scheme  \cite{fdh} is used, the term 
proportional to $\epsilon$ in Eq.(\ref{eq5}) should 
be dropped. 

In Ref.~\cite{Melnikov:2011ta} we have integrated the dipole  
in Eq.(\ref{eq5}) over the {\it restricted}  unresolved phase-space \cite{nagy}, 
drawing extensively 
from the results of Ref.~\cite{Campbell:2004ch}.
We reproduce  this result here for completeness.
We consider the integration of the dipole in Eq.(\ref{eq5})
over the unresolved restricted phase-space 
\be
\begin{split}
& \int \left [{\rm d}g \right ] \; 
\left [ 1-\theta(1-\alpha-z) \theta(y- \alpha y_{\rm max} ) \right ] 
\;D_{g_1 b}^{t} 
= {\cal N} \int \limits_{0}^{1}
{\rm d} z \left ( r^2 + z(1-r^2) \right )^{-\ep}
\\
&  \times \int \limits_{0}^{y_{\rm max}}
{\rm d} y y^{-\ep} (y_{\rm max} - y)^{-\ep}
\;\left [ 1-\theta(1-\alpha-z) \theta(y- \alpha y_{\rm max} ) \right ] 
D_{g_1 b}^{t}.
\end{split}
\label{eq6}
\ee
where 
\be
y_{\rm max} = \frac{(1+r)^2z(1-z)}{z+r^2(1-z)},\;\;\;\;\;\;\;
{\cal N} = \frac{(1-r)^2}{16\pi^2} m_t^{2-2\ep}
\frac{(4\pi)^\ep}{\Gamma(1-\ep)}
\left (\frac{1+r}{1-r}   \right )^{2\ep}.
\ee
We find the following result in CDR
\ba
&& \int \left [ {\rm d}g \right ] \; D_{g_1 b}^{t}
\left [ 1-\theta(1-\alpha-z) \theta(y- \alpha y_{\rm max} ) \right ] =
\nonumber \\
&& \frac{\alpha_s}{2\pi}
\frac{(4\pi\mu^2)^\epsilon}{m_t^{2\ep} \Gamma(1-\ep)}
\delta_{\lambda \lambda'},
\left [ \frac{1}{\ep^2}
+\frac{1}{\ep} \left ( \frac{5}{2} - 2\ln(r_1) \right )
+ \frac{27}{4}
+ \frac{1}{2}
\left (
\frac{1}{r_1^2} - \frac{8}{r_1} + 7 \right ) \ln r^2
\right.
\nonumber \\
&&  \left.
+ \frac{1}{2r_1}
+ 2 {\rm Li}_2(r_1) - \frac{5\pi^2}{6} - 5 \ln(r_1) +
2 \ln^2(r_1) 
\right.
\nonumber \\
&& \left.
- 2 \ln^2 \alpha
- \left ( \frac{7}{2} - 4\alpha
+ \frac{\alpha^2}{2} \right ) \ln \alpha
+\frac{2(1-\alpha)r^2}{r_1}
\ln \left ( \frac{1-r_1}{1-\alpha r_1} \right )
\right ],
\ea
{with $r_1 = 1-r^2$}.

It remains to construct the gluon-gluon dipole of the final-initial type 
for decay kinematics.
 In variance with the gluon-quark  dipole just considered, the gluon-gluon 
dipole contains non-trivial spin correlations. 
We will use the parametrization of the 
unresolved phase-space that we just discussed 
with an obvious modification of the momentum 
${\tilde p}_W$; for the gluon-gluon dipole, it is given by 
${\tilde p}_W = p_W + p_b$.  To derive the gluon-gluon dipole, 
we consider the limit of the $ 0 \to \bar t b g_1 g_2 W$  amplitude
squared  
when two gluons become collinear. The result reads 
\be
|{\cal M}|^2 \to {\cal M}^*_{\mu} P^{gg}_{\mu \nu} {\cal M}^\nu,
\ee
where 
\be
P_{\mu \nu}^{gg} \sim \left [ -g_{\mu \nu} 
\left ( \frac{\xi}{1-\xi} + \frac{1-\xi}{\xi} \right ) 
- 2(1-\ep)\xi (1-\xi) \frac{k_\perp^\mu k_\perp^\nu}{k_\perp^2}
\right ] 
\label{sdsplit}
\ee
is the spin-dependent splitting function. 
In Eq.(\ref{sdsplit}), $\xi$ and $k_\perp^\mu$ are defined as 
\be
p_{g_1}^{\mu} = (1-\xi) p^\mu + k_\perp^\mu - \frac{k_\perp^2 n^\nu}{(1-\xi)(2pn)},
\;\;\;\;
 p_{g_2}^\mu
= \xi p^\mu - k_\perp^\mu - \frac{k_\perp^2 n^\nu}{\xi (2pn)}, 
\ee
where the light-like vector $p$ defines the collinear direction 
and another light-like vector $n_\mu$ is auxiliary. 
We can now use the relations between gluon momenta
\be
k_\perp^\mu \approx a_\mu = \xi p_{g_1}^\mu - (1-\xi) p_{g_2}^\mu, 
\;\;\;\; 2 p_{g_1}  p_{g_2} = - \frac{k_\perp^2}{\xi(1-\xi)},
\ee
to write 
\be
P_{\mu \nu}^{g g}   
\sim \left [ - g_{\mu \nu} \left ( \frac{\xi}{1-\xi}
+ \frac{1-\xi}{\xi} \right )+ (1-\epsilon) \frac{a_\mu a_\nu }{
( p_{g_1} p_{g_2})}
\right ].
\ee
To construct the dipoles, we split this expression into two terms
\be
\frac{P_{\mu \nu}^{gg}}{2 p_{g_1} p_{g_2}} \sim  D^{1,2}_{\mu \nu} 
+ D^{2,1}_{\mu \nu} ,
\ee
where 
\be
D_{\mu \nu}^{1,2} = \frac{1}{2p_{g_1} p_{g_2}} 
\left \{ -\frac{\xi g_{\mu \nu}}{(1-\xi)} 
+ \frac{1-\ep}{2} \frac{a^\mu a^\nu}{(p_{g_1}  p_{g_2})}
\right \}  
\label{eq_d12}
\ee
and $D_{\mu \nu}^{2,1}$ is given by Eq.(\ref{eq_d12}) with $\xi \to 1-\xi$.
We would like to rewrite this equation in such a way that the integration 
over the unresolved phase-space becomes straightforward. To this end, 
we express  Eq.(\ref{eq_d12}) in terms of the variables $z$ and $y$ and 
momentum  of the top quark $p_t$ and ${\tilde p}_W$. 
Because 
\be
\frac{(p_t  p_{g_1})}{(p_t  p_{g_2})} 
= \frac{1-\xi}{\xi},
\ee
we can identify $\xi$ with the variable $z$ in Eq.(\ref{eq4}).
 It remains to modify 
the spin-correlation part of Eq.(\ref{eq_d12}) and write it in appropriate 
variables.   We note that such modifications can be arbitrary 
provided that   the  original form of the spin-correlation part of the 
dipole  is recovered 
in the limit when  $p_{g_1}$ and $p_{g_2}$ become collinear.
We do that by writing 
\be
 a^\mu \to \pi^\mu =  \left ( g^{\mu \nu} 
- \frac{p_t^\mu \tilde p_{12}^\nu + p_t^\nu {\tilde p}_{12}^\mu}{p_t  
{\tilde p}_{12}} \right ) {a_\nu}.
\label{eq_def_pi}
\ee
In Eq.(\ref{eq_def_pi}),  the momentum 
${\tilde p}_{12}$ is the light-like vector 
given by  ${\tilde p}_{12}  =  p_t - \Lambda {\tilde p}_W$, where 
$\Lambda$ is the Lorentz transformation constructed explicitly 
in Ref.~\cite{Campbell:2004ch}.  The reduced matrix element 
that describes the decay process $t \to W + b + g$ is 
then evaluated for $p_t, \Lambda {\tilde p}_W$,
 and ${\tilde p}_{12}$, where $\Lambda {\tilde p}_W$  
is then split into the $W$ momentum and the $b$-quark  momentum. 
We note that the projection 
operator introduced in Eq.(\ref{eq_def_pi}) ensures that
$\pi_\mu$  is transverse to 
${\tilde p}_{12}$. As we show below, this feature 
simplifies the integration over the unresolved phase-space considerably.
It is straightforward  to check that in the collinear  $(y \to 0)$
limit,  $\pi_\mu \to a_\mu$. Hence, to construct a suitable dipole, 
we can simply substitute $\pi_\mu$ for $a_\mu$ in Eq.(\ref{eq_d12}).
Note also that we are allowed to multiply the spin-correlation part 
in Eq.(\ref{eq_d12}) by an arbitrary function $f(y,z)$
provided that it is free of singularities and that 
it is normalized in such a way that $f(0,z) = 1$. We choose this function 
to be 
\be
f(y,z) = \frac{4}{m_t^{{4}}} \frac{(p_t {\tilde p}_{ W})^2 - r^2 m_t^4}{(1-r^2)^2},
\ee
to simplify the calculation of the integrated dipole with $\alpha$-dependence.  As the very last
step, we add one more term to the dipole, to account for soft singularities 
that appear when a gluon is emitted from the top quark in the initial 
state.  We are finally in the position to 
write down the  $gg$ final-initial dipole. In the CDR scheme, the 
result reads 
\be
D_{g_1,g_2}^{\mu \nu,t} = 4 \pi \alpha_s \mu^{2\ep} { {1\over 2p_{g_1}p_{g_2}}}
\left [ - g^{\mu \nu} \left ( \frac{z}{1-z}  - \frac{m_t^2}{4}
\frac{2 p_{g_1}  p_{g_2}}{( p_t  p_{g_1}  )^2} \right )
+ \frac{(1-\epsilon)  \pi^\mu \pi^\nu }{2 p_{g_1}  p_{g_2}}
f(y,z)
\right ].
\label{eq_gg_dip}
\ee
The various quantities that appear in Eq.(\ref{eq_gg_dip}) are 
\be
\begin{split} 
& \pi^\mu = \frac{1}{p_t   {\tilde p}_{12}}
\left ( ( p_t   {\tilde p}_{12} ) a^\mu 
- p_t^\mu ({\tilde p}_{12}  a ) \right ),\;\;\;\;
a^\mu = \frac{2}{m_t^2(1-r^2)} 
\left [ (p_t p_{g_2}) p_{g_1}^{\mu} 
- \left ( p_t  p_{g_1} \right ) p_{g_2}^\mu \right ] ,
\\
& p_t  p_{g_1} = \frac{m_t^2}{2} (1 - r^2) (1-z),
\;\;\;\; p_{g_1}  p_{g_2} = \frac{m_t^2}{2}(1-r)^{{2}} y,\;\;\;
p_t  {\tilde p}_{12} = \frac{m_t^2 (1 -r^2)}{2},
\end{split} 
\ee
with $ r^2  = (p_W + p_b)^2/m_t^2$.

To integrate the dipole in Eq.(\ref{eq_gg_dip}) over the unresolved 
phase-space,  we make use of the results presented in 
Ref.~\cite{Campbell:2004ch}.  It is straightforward 
to integrate the part proportional to the metric tensor. 
Integration of the spin-correlation part is more involved but it 
can be  simplified because vector $\pi^\mu$ is 
orthogonal to the light-like vector 
${\tilde p}_{12}^\mu$.  This allows us to  write 
\be
\left \langle \frac{\pi^\mu \pi^{\nu} }{(2 p_1  p_2)^2} \right \rangle_{y,z} 
= A_1 \left ( - g^{\mu \nu} + 
\frac{p_t^\mu {\tilde p}_{12}^\nu 
+ {\tilde p}_{12}^\mu p_t^\nu }{p_t  {\tilde p}_{12}} \right )  
+ A_2 {\tilde p}_{12}^\mu {\tilde p}_{12}^\nu,
\label{eq_tens}
\ee
where $\langle... \rangle_{y,z}$ denotes the integration over $y$ and $z$ 
as in Eq.(\ref{eq6}). The term proportional to 
$A_2$ can be dropped since it  gets Lorentz-contracted with the 
product of on-shell matrix elements that vanish when contracted 
with  ${\tilde p}_{12}$.  Hence, we only need 
to compute $A_1$, which we easily obtain by contracting the 
left hand side of the above formula 
with the metric tensor. By the same argument, once $A_1$ is obtained, 
we can drop terms proportional to $\tilde p_{12}^\mu$ in tensorial 
structure that is multiplied by $A_1$ in Eq.(\ref{eq_tens}). Therefore, 
we can write the result of the integration of $D_{g_1,g_2}^{\mu \nu, t}$
over unresolved phase-space  
as proportional to the metric tensor. 

We now present the result for the integrated  final-initial 
$gg$ dipole in the CDR scheme for decay kinematics, including
its full $\alpha$-dependence. The integrated 
dipole reads
\be
\begin{split}
 \int & \left [ {\rm d}g \right ] \; D_{g_1 g_2}^{\mu \nu,t}
\left [ 1-\theta(1-\alpha-z) \theta(y- \alpha y_{\rm max} ) \right ] = 
\frac{\alpha_s}{2\pi} \frac{ ( 4 \pi \mu^2 )^{\ep}}{m_t^{2\ep} \Gamma(1-\ep) }
 \; g_{\mu \nu} \times \Bigg [ 
\\
& \frac{1}{2\ep^2} + \frac{17-12 \log r_1}{12\ep}
- \frac{5 \pi^2}{12}
- \log^2 \alpha - \frac{\left(1-\alpha \right ) \left ( 
23 - \alpha + 2 \alpha^2 \right )}{12} \log \alpha
+ \log^2 r_1 
\\
& - \frac{17}{6} \log r_1 
- \frac{r^2 \log r}{6 r_1^5}
\Big [ 6\alpha^3 (1 - r_1) (-2 + r_1) - 3 \alpha^2 (1 - r_1) (-6 + 5 r_1) \\
&+  12 \alpha r_1 (r^2 + r_1^3) + r_1^2 (2 + r_1 (-1 + 11 r_1))
 \Big ]
+ \frac{(1-\alpha) r^2 \log( 1- \alpha r_1)}{4 r_1^5}
\Big [ 
\\
& 
(- 2 \alpha^2 (1 - r_1) (-2 + r_1) + \alpha (-2 + (5 - 3 r_1) r_1) -2 + r_1 + r_1^2 - 4 r_1^4) 
\Big ] + {\rm Li}_2(r_1)
\\
& 
 - \frac{1}{240 r_1^4 ( 1- \alpha r_1)} \Big [
 - 8 \alpha^9 r_1^5
 -6 \alpha^8 r_1^4 \left (2 - 7 r_1 \right )
 - \alpha^7 r_1^3 (20 - 68 r_1 + 115 r_1^2)
\\
& 
+ \alpha^6 r_1^2 ( -40 +130 r_1 - 165 r_1^2 + 216 r_1^3) 
- \alpha^5 r_1 ( 120 - 360 r_1 + 410 r_1^2  -234 r_1^3 + 305 r_1^4)
\\
& 
+ \alpha^4  ( 240 - 180 r_1 - 510 r_1^2 + 650 r_1^3 - 195 r_1^4 
+ 278 r_1^5)\\
&
- \alpha^3 (600 - 1140 r_1 + 280 r_1^2 + 460 r_1^2 - 92 r_1^4 + 97 r_1^5) \\
&+ \alpha^2 ( 360 - 1140 r_1 + 900 r_1^2 + 50 r_1^3 - 63 r_1^4 - 40 r_1^5 )
\\
& + 10 \alpha \left ( 12 + 6 r_1 - 36 r_1^2 + 10 r_1^3 + 8 r_1^4 
+ 91 r_1^5 \right )  
+ 10 r_1^2 ( 4 r^2  - 91r_1^2) 
\Big ] \Bigg ].
\end{split} 
\label{eq19}
\ee
The integrated dipole given in Eq.(\ref{eq19}) is the final  ingredient  
we need to treat the real emission contributions to radiative decays 
of top quarks.

\section{Phenomenological results} 

In this Section we present  phenomenological results for the Tevatron 
($\sqrt{s} = 1.96~{\rm TeV}$) 
and the LHC ($\sqrt{s} = 7~{\rm TeV}$).
We choose $m_t = 172~{\rm GeV}$
for the top quark mass and $m_W = 80.419~{\rm GeV}$
for the $W$-boson mass. We employ
MSTW2008  parton distribution functions  \cite{Martin:2009iq}
and use the corresponding values of $\alpha_s$ at leading 
and next-to-leading order.
The couplings of the $W$-boson to fermions are obtained from
the Fermi constant $G_\mathrm{F} = 1.16639 \cdot 10^{-5} \,{\rm GeV}^{-2}$.
Since we work in the narrow
width approximation, our results are
inversely proportional to the top quark and the $W$-boson
widths, $\sigma \sim \Gamma_t^{-2} \Gamma_W^{-2}$. These decay widths
are evaluated at leading and next-to-leading order in the strong 
coupling constant, for LO and NLO
cross-sections, respectively. For reference, we give the results for 
the widths
\be
\begin{split}
& \Gamma_t^{\rm LO} = 1.4653~{\rm GeV},~~~~~\Gamma_t^{\rm NLO} = 1.3375~{\rm GeV},
\\
& \Gamma_{W}^{\rm LO} = 2.0481~{\rm GeV},~~~~
\Gamma_{W}^{\rm NLO} = 2.1195~{\rm GeV}.
\end{split}
\ee
The shown NLO results for the widths are computed 
with the renormalization 
scale $\mu = m_t$.  We note that the use of NLO expressions for the
widths increases the NLO cross-sections by about ten percent.

We begin with the discussion of the Tevatron results.
We consider $t \bar t$ production in the lepton + jets 
channel so that our leading order cross-section 
contains five jets.
The lepton transverse momentum and the missing
energy in the event are required to satisfy $p_{\perp,l}> 20~{\rm GeV}$ and 
$E_\perp^{\rm miss} > 20~{\rm GeV}$.  Jets are defined 
according to the $k_\perp$-jet algorithm 
\cite{Catani:1993hr} with $\Delta R = 0.5$.
The jet transverse momenta   are required to be
larger than
$p_{\perp,j} > 20~{\rm GeV}$. Both leptons and  
jets must be central $|y_l| < 2,\;|y_j| < 2$.
To better discriminate against the background,
we require  an additional cut on the transverse energy in
the event $H_\perp = \sum_{j} p_{\perp,j} +
p_{\perp,e} + E_{\perp}^{\rm miss} > 220~{\rm GeV}$.
We present results below for a single lepton generation. Hadronic decays
of $W$-bosons to first two quark generations are included  and the CKM
matrix is set to the identity matrix.

The cross-sections for 
$p \bar p \to b W^+(e^+ \nu_e)  \;  \bar b W^-(jj) +j$ production
at the Tevatron at leading and next-to-leading order 
in perturbative QCD, subject to the above cuts, read 
\be
\sigma_{\rm LO} = 75.29^{+ 49.2}_{-27.4}~{\rm fb},\;\;\;\;
 \sigma_{\rm NLO} = 78.9^{-5.6}_{-5.6}~{\rm fb}.
\label{eq21}
\ee
In Eq.(\ref{eq21}), the central value refers to renormalization and 
factorization scales set to $\mu = m_t$ and the upper (lower) value to 
$\mu = m_t/2$ and $\mu = 2m_t$, respectively.  We observe a dramatic reduction 
in dependence on unphysical scales if NLO QCD corrections are included.

\setlength{\abovecaptionskip}{-2ex}
\begin{figure}[t]
 \begin{center}
 \includegraphics[scale=1]{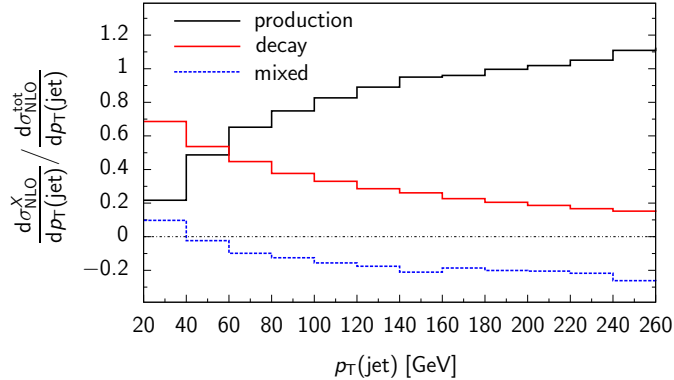}
\end{center} 
\caption{
Fractions of events when the leading (non-$b$) jet at the 
Tevatron  comes from $t \bar t j$ production, the decay
$ t \to W b j$ or mixed processes, as a function of jet transverse momentum. Note the 
sign of the mixed contribution and the cancellation between decay and mixed 
mechanisms at high transverse momentum. Renormalization and factorization scales are set to $\mu = m_t$.}
\label{fig0}
\end{figure}

It is interesting to understand how jet radiation  in the production 
and jet radiation in the decay contribute to cross-sections shown in 
Eq.(\ref{eq21}).  To answer this question, we present separate 
cross-sections for production and decay processes as well as mixed 
contributions, as defined in Eq.(\ref{eq20}).
For factorization and
renormalization scales set to $\mu = m_t$ we find
\be
\begin{split}
& \sigma_{\rm LO} = 46.33 \left ( {\rm Pr} \right ) 
+ 28.96 \left ( {\rm Dec} \right) 
= 75.29~{\rm fb},
\\
& \sigma_{\rm NLO} = 
47.7  \left ( {\rm Pr} \right ) 
+36.7 \left ( {\rm Dec} \right ) - 5.5\left ( {\rm Mix} \right ) = 78.9~{\rm fb}. 
\end{split} 
\label{eq22}
\ee
This result is  interesting because it shows that, with our choice 
of selection criteria,  in only sixty percent of all events that contain 
a $t \bar t$ pair and a jet, the jet can be associated with the production 
process; in the remaining forty percent of events, jets come from top quark 
decays.
These fractions are stable against NLO QCD corrections, but the reason 
for that stability is peculiar. Indeed, it follows from Eq.(\ref{eq22}) 
that the NLO QCD corrections to the production process are relatively 
small ($K = 1.03$) 
while QCD corrections to the decay process are quite large
($K = 1.37$).  There is, however, a significant 
{\it negative} contribution from the ``mixed'' corrections.
As described around Eq.(\ref{eq20}), this contribution arises
from single jet emission in the production convoluted with 
single jet emission in the decay and the corresponding virtual corrections.
Because of this cancellation between decay and mixed contributions,
a relatively small correction to  jet radiation in 
top quark decays remains. Thus,
an estimate of the NLO cross-section that employs 
the exact leading-order cross-section as in Eq.(\ref{eq21}) and the $K$-factor
for the production process $K = 1.03$ gives $\sigma_{\rm LO} \times K = 77.54$, 
which is in good agreement with the full NLO result ($\mu = m_t$) in
Eqs.(\ref{eq21},\ref{eq22}).
However, this cancellation   seems accidental to us.    
In spite of the proximity of the two numbers
for the $t \bar t j$ production at the Tevatron, we were unable  
to come 
up with a convincing and general argument that ensures that $K$-factors for
the production and decay processes are always similar. 
In fact, the importance of mixed and decay contributions strongly
depends on the kinematic variables.
For illustration
we show production, decay and mixed contributions 
as the function of the transverse momentum of the leading non-$b$ jet in Fig.~\ref{fig0}.
At low $p_\perp^{\rm jet} \lesssim 60\, \rm{GeV}$, jet radiation  in top quark 
decays is the largest ($\sim 60\%$) contribution to  the cross section.
As expected, at larger $p_\perp^{\rm jet}$, the  jet is predominantly 
emitted in the $t\bar t $ production.
The mixed contribution is positive at small jet momenta but changes sign
at moderate $p_\perp^{\rm jet}$ and 
cancels the contribution due to jet radiation  in decay at large
$p_\perp^{\rm jet}$.  The situation  appears to be quite complex and 
observable-dependent.  We can therefore anticipate -- and
we will see this explicitly in the context of the LHC discussion --
that calculations  without accounting for jet radiation
in the decays of top quarks can lead to  misleading results.

\begin{figure}[!t]
 \begin{center}
 \includegraphics[width=8cm,height=4.9cm]{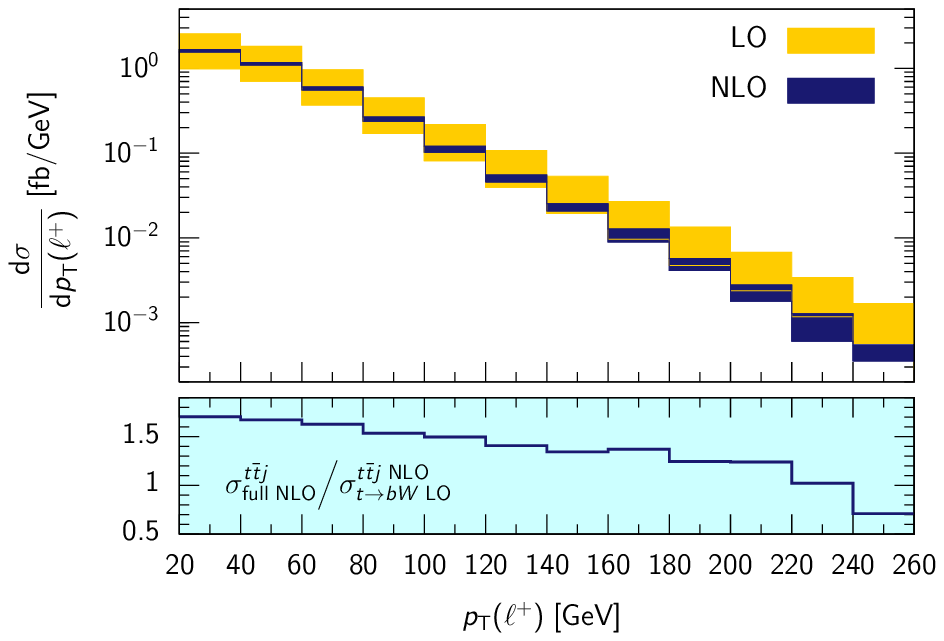}~~~~~
 \includegraphics[width=8cm,height=5cm]{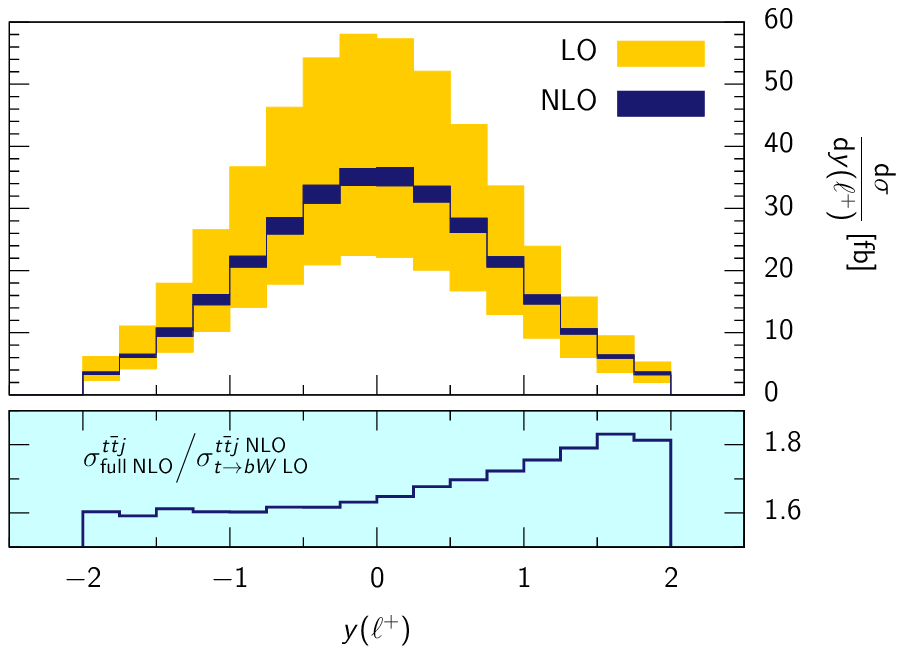} \\[2ex]
 \includegraphics[width=8cm,height=5cm]{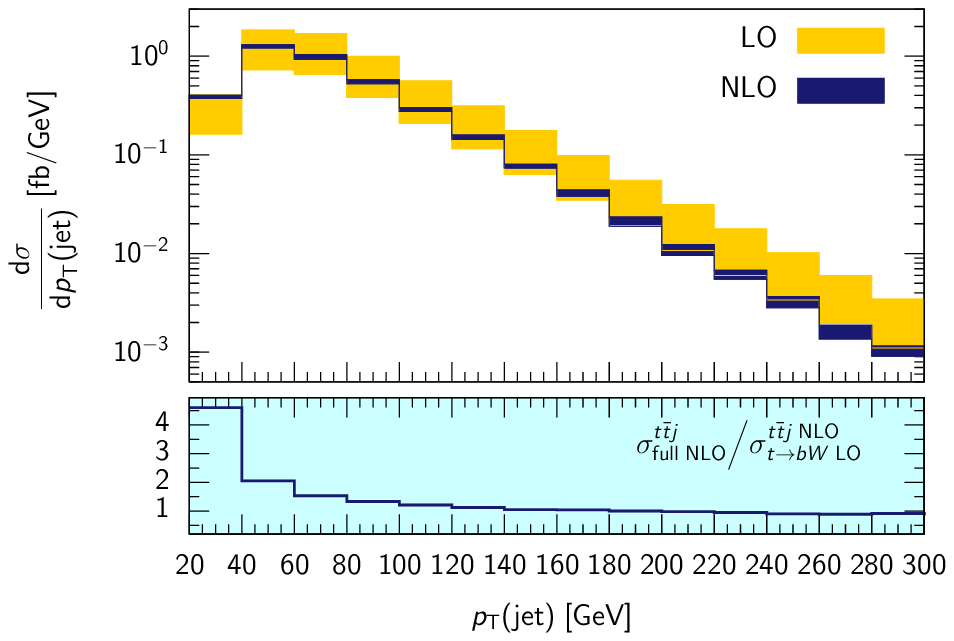}~~~~~
 \includegraphics[width=8cm,height=5cm]{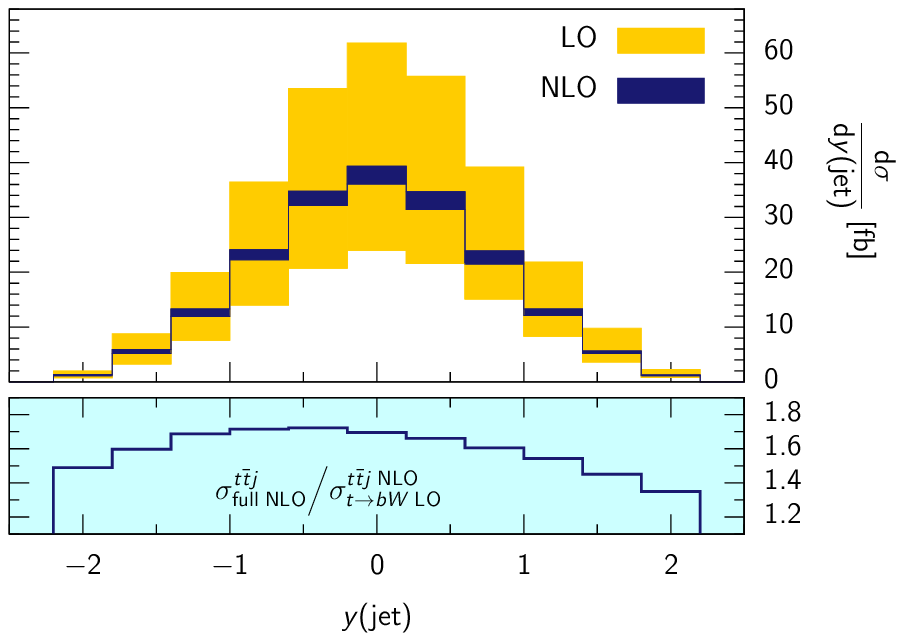}
\end{center} 
\caption{
Distributions of the lepton transverse momentum, the lepton rapidity, the transverse momentum and the rapidity of the
hardest jet for $\ttbj$ production at the Tevatron at leading and next-to-leading order in perturbative QCD.
The bands correspond to the variation of
renormalization and factorization scales
in the interval $m_t/2 < \mu <  2m_t$.
Results with hard jet emission in the \textit{production} stage only followed by leading
order decays $t \to W+b$ are compared to full NLO results in lower panes.
}
\label{fig1}
\end{figure}

\begin{figure}[t]
 \begin{center}
\includegraphics[width=8.3cm,height=5cm]{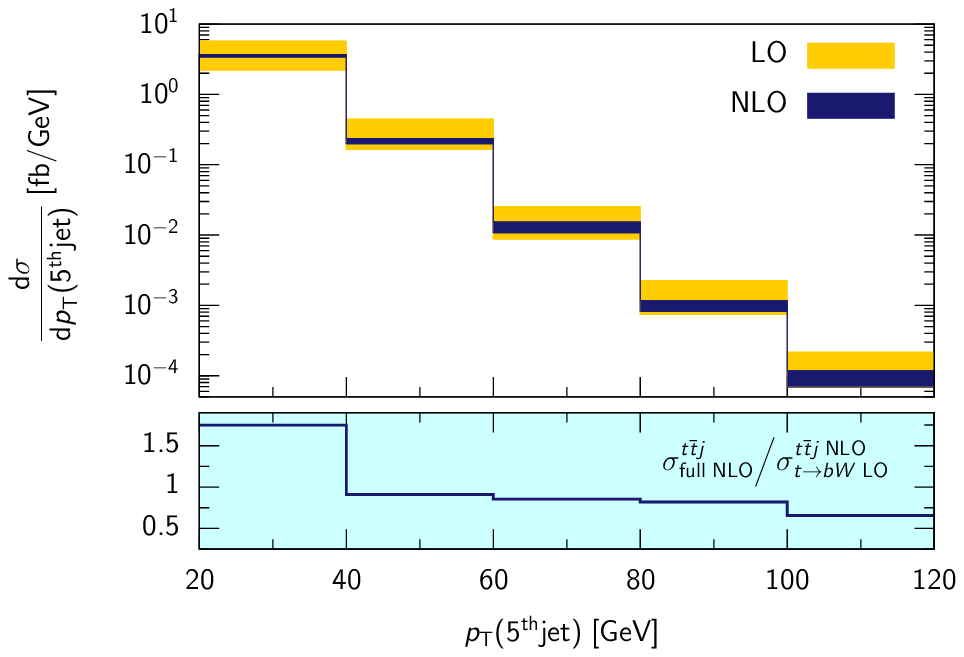}~~
\includegraphics[width=7.7cm,height=4.9cm]{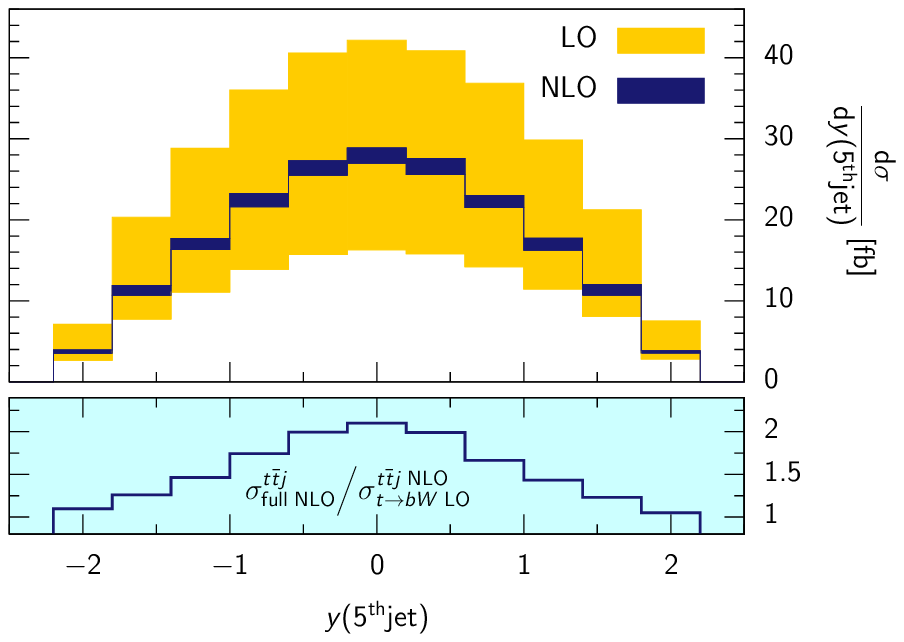}~~~~~~~~~~~~~~~~\\[2ex]
~~\includegraphics[width=8cm,height=5cm]{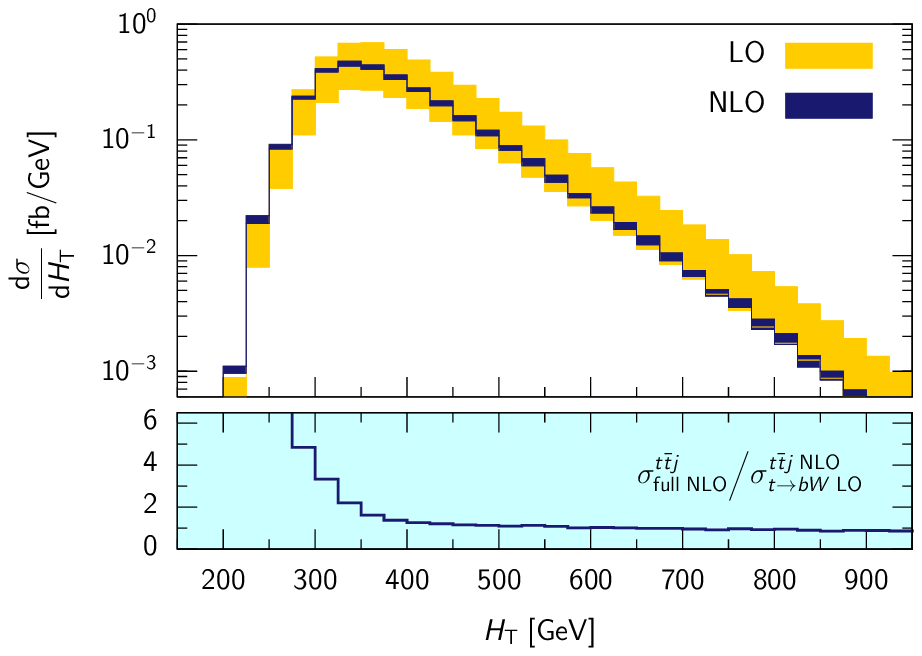}~~~
\includegraphics[width=8cm,height=4.85cm]{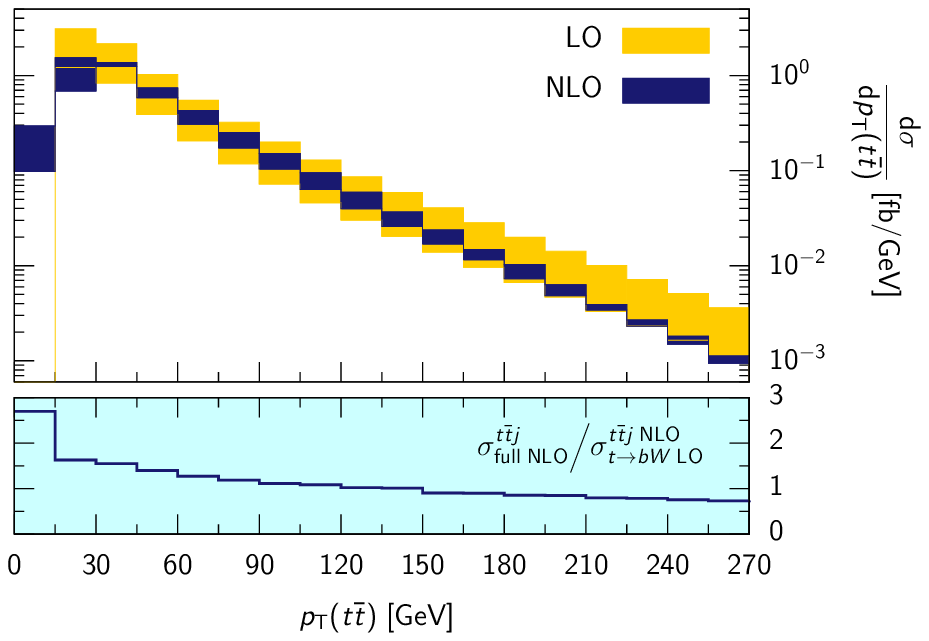}
\end{center}
\caption{ 
Distributions of the transverse momentum and the rapidity of the
5th hardest jet, the transverse energy $H_{\rm T}$ and the transverse momentum of the $t\bar{t}$ pair for $\ttbj$ production
at the Tevatron at leading and next-to-leading order in perturbative QCD.
The bands correspond to the variation of
renormalization and factorization scales
in the interval $m_t/2 < \mu <  2m_t$.
Results with hard jet emission in the \textit{production} stage only followed by leading
order decays $t \to W+b$ are compared to full NLO results in lower panes.
}
\label{fig2}
\end{figure}

Various kinematic distributions at the  Tevatron are shown in 
Figs.~\ref{fig1} and \ref{fig2}.  For all kinematic distributions
we find a significantly reduced dependence on the choice of the factorization and the  renormalization scales as well as 
shape changes in kinematic tails of some distributions.
The impact of QCD radiation in top quark decays is illustrated in the 
lower panes of each plot,
where ratios of full
NLO cross-section and the NLO $t \bar t j$ production cross-section 
followed by the leading order decays of top quarks are  shown. 
In general, these plots confirm the expectation that QCD radiation 
in top quark decays mostly affects spectra at low transverse momenta. 
But there are interesting exceptions where the impact of radiation
in the decay is more pronounced. 
In particular, we find fairly uniform enhancement 
of transverse momenta and rapidity distributions of the 
charged lepton as well as
the rapidity of the hardest jet (Fig.~\ref{fig1}).  The decay contribution
to the rapidity distribution of a lepton is asymmetric; it appears 
to be  more important at large positive rapidities. 
However, the full NLO distribution does not show  significant asymmetry in lepton rapidity. 

In Fig.~\ref{fig2} we show distributions of the transverse momentum
and rapidity of the 5th hardest jet,
the total transverse energy in the event $H_\perp$ 
and the transverse momentum of the $t \bar t$ pair. 
All these distributions receive non-uniform enhancements from jet 
radiation in top quark decays.
In particular, $H_\perp$ and $p_\perp({\rm 5th~jet})$ distributions are 
strongly enhanced  at  low values of $H_\perp$ and $p_\perp$, 
where relatively soft radiation in  top quark decays 
dominates.
Also, the rapidity distribution of the 5th hardest jet receives
strong enhancement at central rapidities which is a 
consequence of the fact that top quark decay products 
are produced mostly at small rapidities.
We note that similar shape changes were recently 
observed in the context of studying $p \bar{p} \to t \bar t j$ 
within the  parton shower approximation in  Ref.~\cite{Kardos:2011qa}.
Note, however, that the  cross-section computed in 
Ref.~\cite{Kardos:2011qa} seems closer to the contribution that
 we identify as  ``jet radiation in production''.  While -- as we just saw --
such a result underestimates the cross-section, it is probably consistent 
with   the fact that decays in Ref.~\cite{Kardos:2011qa} are treated in 
the parton shower approximation which by construction conserves the overall
probability and does not change normalization.

\begin{figure}[t!]
\begin{center}
\includegraphics[width=7.5cm,height=5cm]{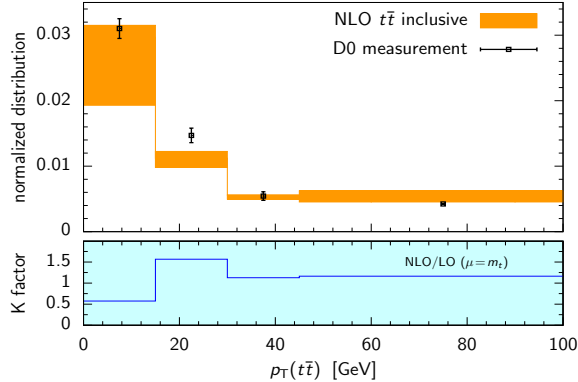}
\end{center}
\caption{ 
Comparison of D0 measurement with theoretical NLO QCD prediction. 
The data points are obtained from Ref.~\cite{Abazov:2011rq} after background subtraction.
The bands correspond to the variation of
renormalization and factorization scales
in the interval $m_t/2 < \mu <  2m_t$. 
The experimental distribution and the $\mu = m_t$ theoretical distribution
 are normalized in such a way that 
their integrals equal to one.}
\label{fig2a}
\end{figure}

We also consider the distribution in the transverse momentum 
of the $t \bar t$ pair in Fig.~\ref{fig2}.
This kinematic distribution is 
particularly interesting 
because recent results by the D$0$ collaboration 
\cite{Abazov:2011rq} show a disagreement between 
predictions of MC@NLO \cite{mcnlo} and data at low transverse momenta.
Since we deal with top quark decay products rather than with stable top quarks, we need to define what is 
meant by the $t \bar t$ transverse momentum.
To this end, we imagine that the reconstruction proceeds
by finding  two non-$b$ jets whose invariant mass is closest to $M_W$ and 
then combining the transverse momenta of these two jets, two $b$-jets, 
the lepton transverse momentum and the missing transverse momentum, 
to obtain the transverse momentum of the $t \bar t$ pair.  
We find that the transverse momentum distribution of the 
$t \bar t$ pair  is affected by the radiation in the decay non-uniformly -- 
the decay contributions are more important for small values of 
$p_\perp(t \bar t )$.

To further compare the results of our computation with  D0 data \cite{Abazov:2011rq}, 
we combine the $p \bar p \to t \bar t j$ calculation   described above with a $p \bar p  \to
t \bar t$ computation at NLO QCD \cite{ms}. In the $p \bar p \to t \bar t$ computation we impose a jet veto prohibiting 
additional jets with the transverse momentum larger than $20~{\rm GeV}$, 
for  consistency  with the current $t \bar t j$ computation.  We present our 
results\footnote{We note that the kinematic cuts on the final state particles that we use are similar but not identical to the ones used by D0 collaboration.} 
and the 
D0 data \cite{Abazov:2011rq} in Fig.~\ref{fig2a}. The normalization of the $\mu = m_t$ 
NLO  computation is chosen such that the integrals of the two distributions agree. 
In spite of the significant theoretical uncertainty  in the lowest bin, it appears 
that our calculation can well describe  the shape of the $p_\perp(t \bar t)$ distribution 
observed by the D0 collaboration. The lower pane in Fig.~\ref{fig2a} shows that inclusion 
of NLO QCD corrections to $pp \to t \bar t j$ and $pp \to t \bar t$ 
is crucial for achieving the agreement. \\

\begin{figure}[t]
 \begin{center}
 \includegraphics[scale=1]{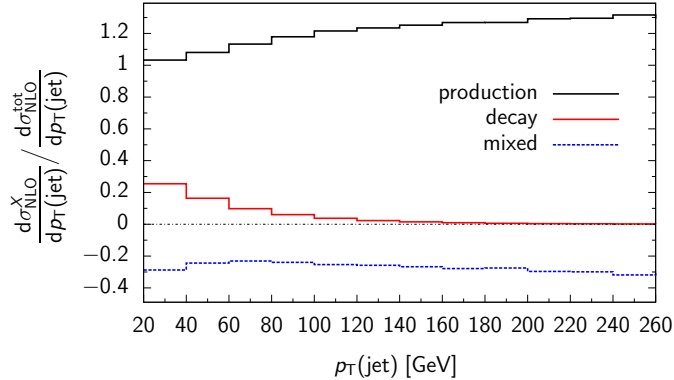}
\end{center}
\caption{
Fractions of events when the leading (non-$b$) jet at the $7~{\rm TeV}$
LHC  comes from $t \bar t j$ production, the decay
$ t \to W b j$ or mixed processes, as a function of jet transverse momentum. Note the
sign of the mixed contribution and the cancellation between decay and mixed
mechanisms at high transverse momentum. Renormalization and factorization scales are set to $\mu = m_t$.}
\label{fig3a}
\end{figure}
We continue with the discussion of $t \bar t j$ production
at the $\sqrt{s} = 7~{\rm TeV}$ LHC.  We imagine that $W$-bosons 
from both $t$ and 
$\bar t$ decays decay leptonically.  
For definiteness, we assume that the top quark decays to a positron
and the antitop quark decays to an electron.  
All generic input parameters that we employ in 
the calculation were already described 
at the beginning of Section III. Specific to the LHC case, we 
require at least three jets, defined by the anti-$k_\perp$ jet algorithm \cite{Cacciari:2008gp} with $\Delta R=0.4$.
All jets have a minimum transverse momentum $p_{\perp,j}>25~{\rm GeV}$ 
and central rapidities $|y_j|<2.5$. Similarly, leptons need to satisfy $p_{\perp,l}>25~{\rm GeV}$  and $|y_l|<2.5$,
and the missing energy in the event $p_{\perp}^{\rm miss}>50~{\rm GeV}$.
\begin{figure}[t]
 \begin{center}
~\includegraphics[width=8cm,height=5cm]{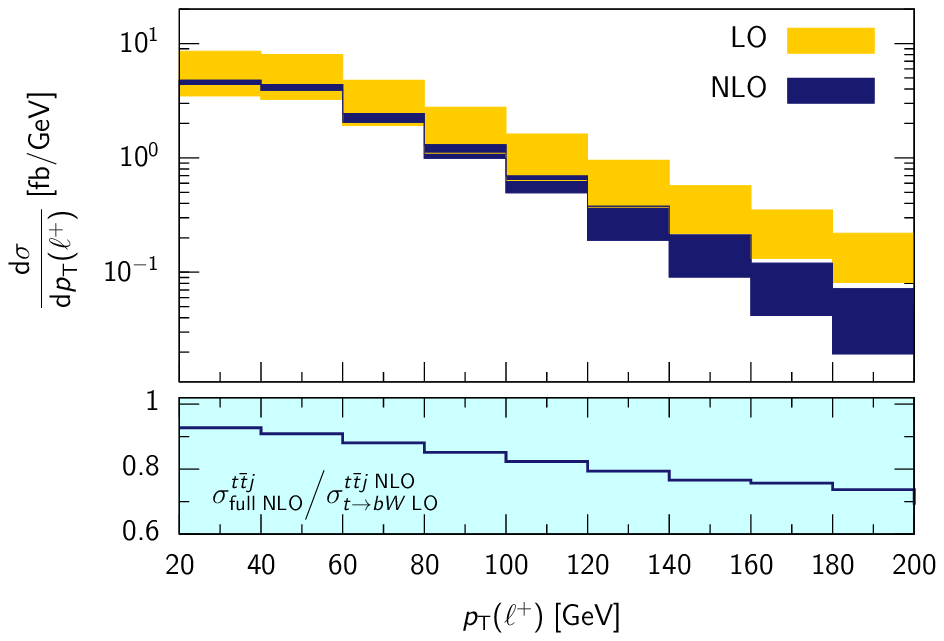}~~
\includegraphics[width=8cm,height=4.85cm]{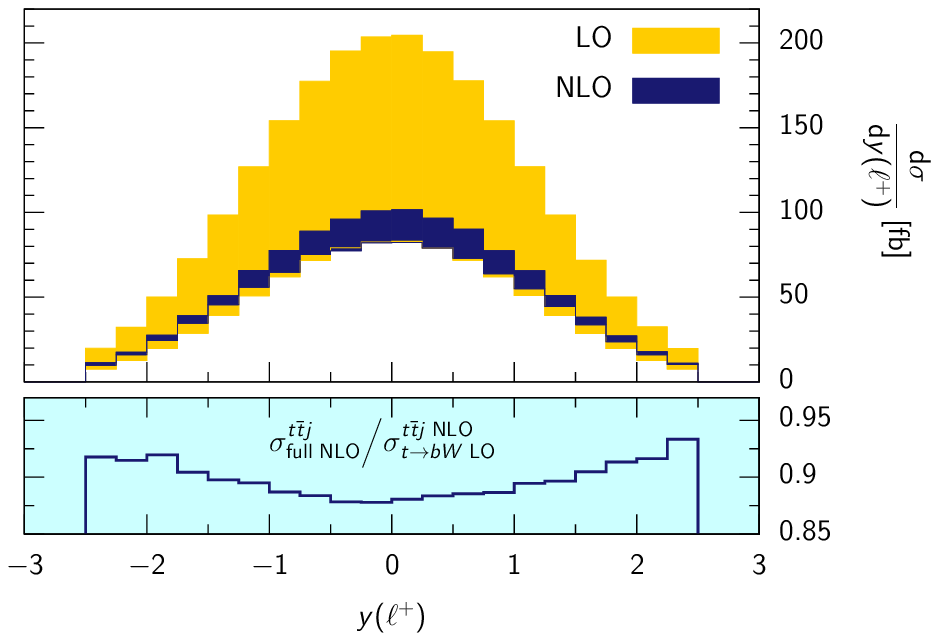}
\\[2ex]
\includegraphics[width=8cm,height=5cm]{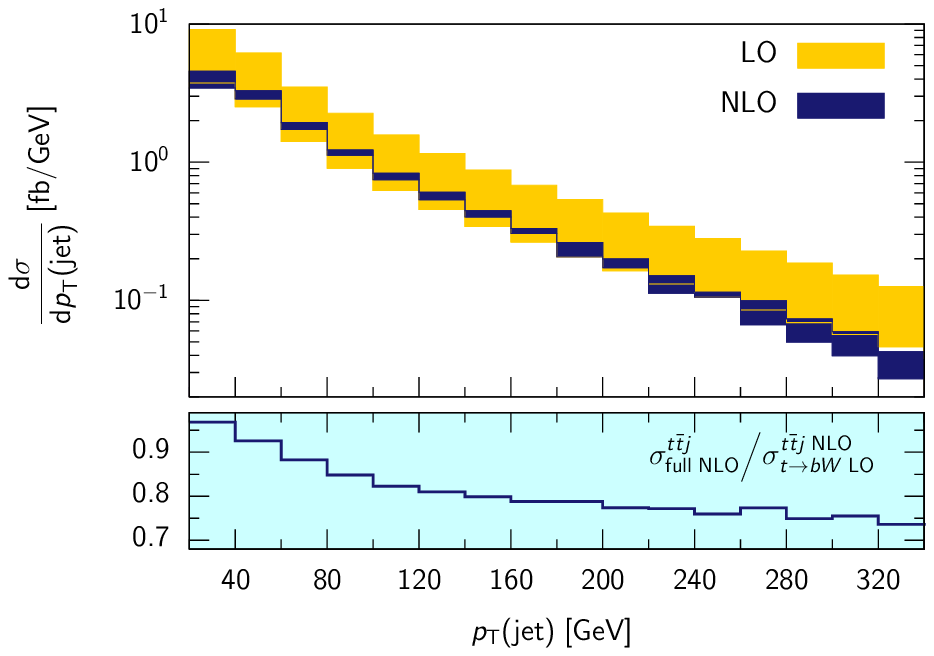}~~~
\includegraphics[width=8cm,height=5cm]{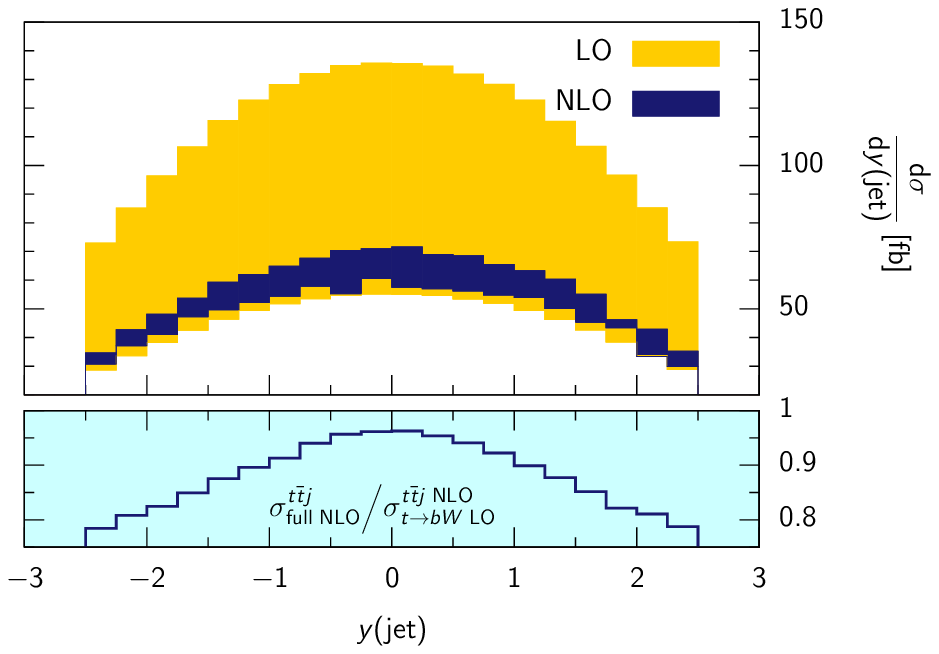}~
\end{center}
\caption{
 Distributions of the lepton transverse momentum, the lepton rapidity, the transverse momentum and the rapidity of the
hardest jet for $\ttbj$ production at the LHC (7 TeV) at leading and next-to-leading order in
perturbative QCD.
The bands correspond to the variation of
renormalization and factorization scales
in the interval $m_t/2 < \mu <  2m_t$.
Results with hard jet emission in the \textit{production} stage only followed by leading
order decays $t \to W+b$ are compared to full NLO results in lower panes.
}
\label{fig3}
\end{figure}
We find the following results 
for leading and next-to-leading order cross-sections 
\be
\sigma_{\rm LO} = 350.3^{+215.0}_{-123.1}~{\rm fb},
\;\;\;
\sigma_{\rm NLO} = 288^{-46}_{-18}~{\rm fb}.
\label{eq25}
\ee
In Eq.(\ref{eq25}), the central value refers to renormalization and 
factorization scales set to $\mu = m_t$ and the upper (lower) value to 
$\mu = m_t/2$ and $\mu = 2m_t$, respectively. 

In case of the LHC, the interplay between radiation in the production 
and  radiation in the decay is very different from the Tevatron.
Since top quark pairs at the LHC are
mostly produced in  gluon annihilation and the collision energy
is high, radiation in the production strongly
dominates over radiation in the decay. We find $(\mu = m_t)$
\be
\begin{split} 
& \sigma_{\rm LO} = 316.9~({\rm Pr}) + 33.4~({\rm Dec}) 
= 350.3~{\rm fb}, \\
& \sigma_{\rm NLO}= 323~({\rm Pr})  + 40.5~({\rm Dec}) - 75.5~({\rm Mix}) 
= 288~{\rm fb}. 
\label{eq26}
\end{split} 
\ee

The three NLO contributions are shown in Fig.~\ref{fig3a}, as a function
of the transverse momentum of the leading non-$b$ jet. The radiation in  the 
decay becomes less and less important as the process becomes harder,  but 
the negative mixed contribution appears to be significant also at high 
$p_\perp$.
Although  radiation in the decay at the 
LHC is less important 
than at the Tevatron, it is peculiar that 
``mixed'' contributions are large and 
negative. 
\begin{figure}[t]
 \begin{center}
\includegraphics[width=7.8cm,height=5cm]{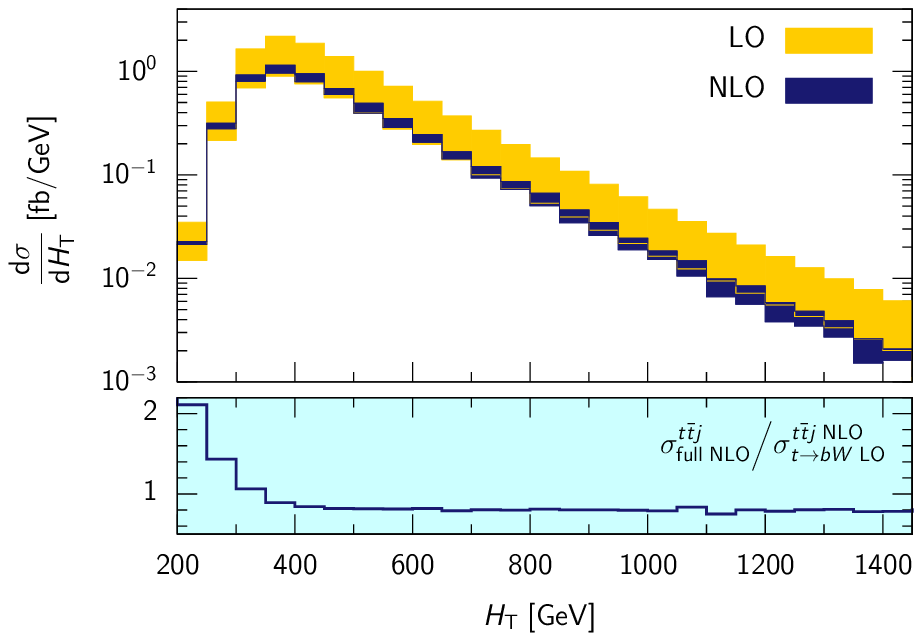}~~~
\includegraphics[width=8cm,height=5cm]{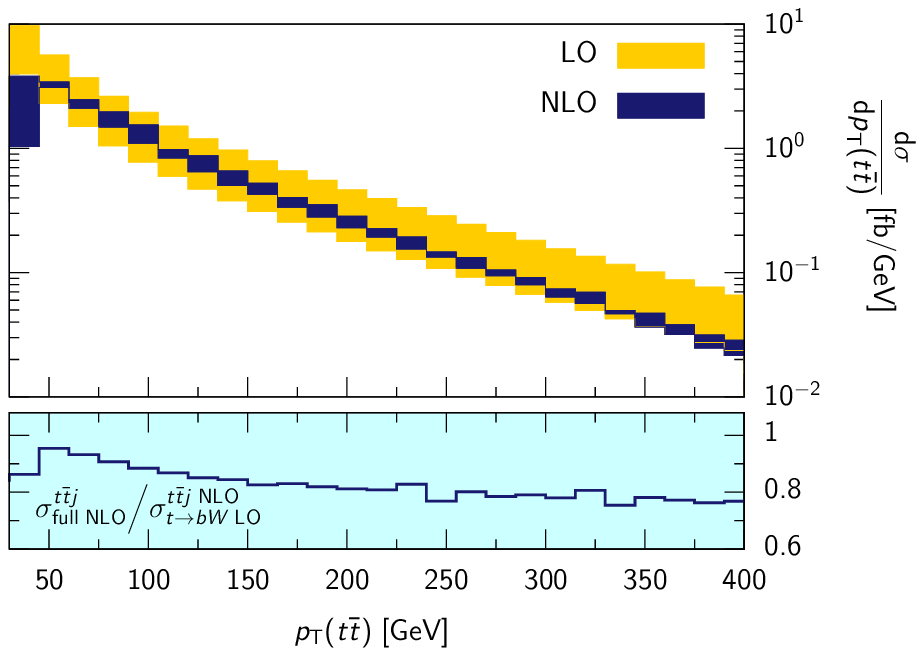}
\\[2ex]
\includegraphics[width=8cm,height=4.9cm]{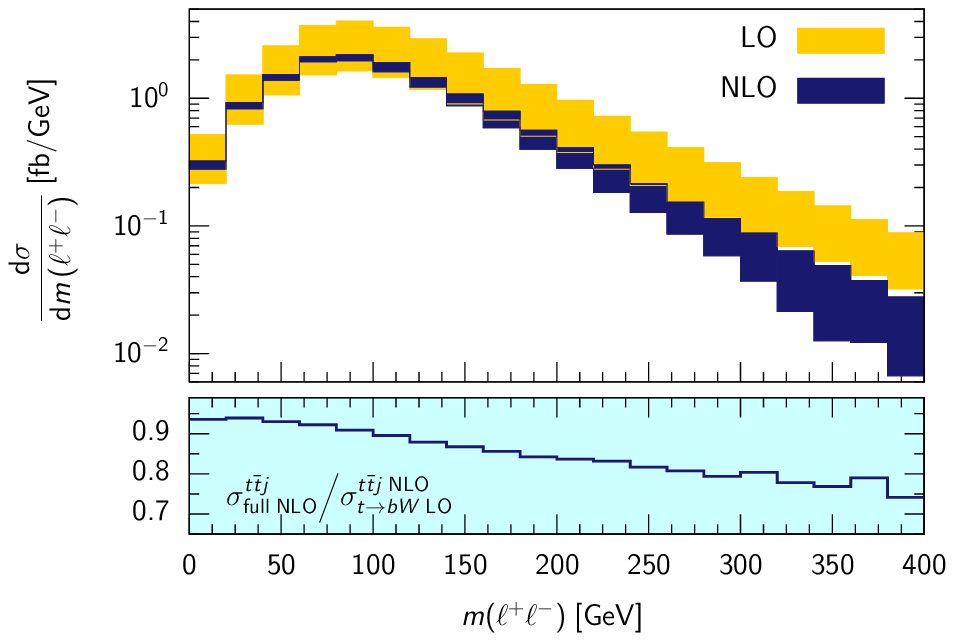}~~
\includegraphics[width=8cm,height=5cm]{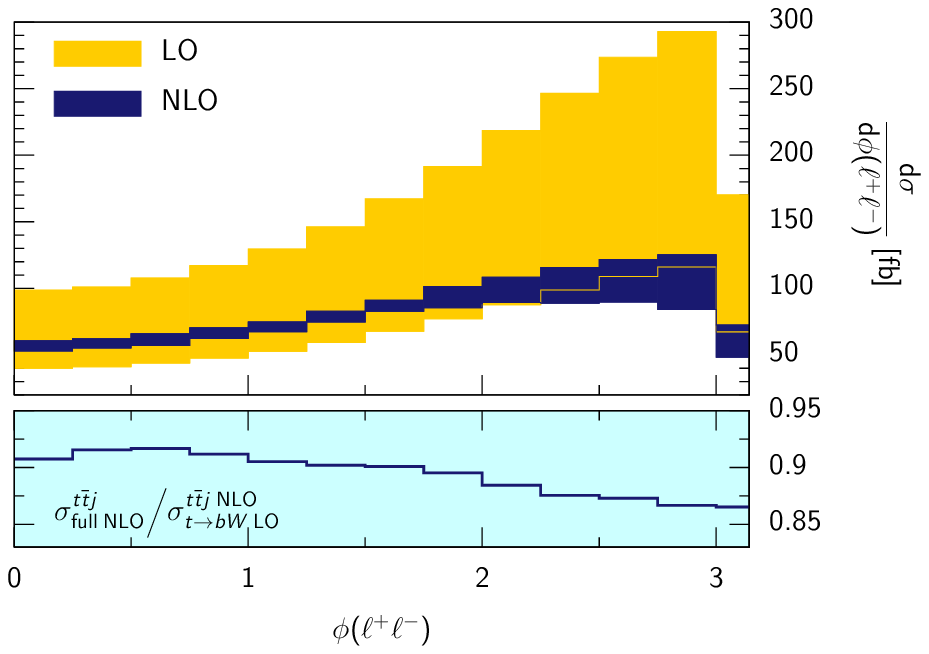}
\end{center}
\caption{
Distributions of the transverse energy $H_{\rm T}$, the transverse momentum of the top quark pair, the
di-lepton invariant mass and the relative azimuthal angle between the leptons
for $\ttbj$ production at the LHC (7 TeV) at leading and next-to-leading order in perturbative QCD.
The bands correspond to the variation of
renormalization and factorization scales
in the interval $m_t/2 < \mu <  2m_t$.
Results with hard jet emission in the \textit{production} stage only followed by leading
order decays $t \to W+b$ are  compared to full NLO results in lower panes.
}
\label{fig4}
\end{figure}

We point out that this may cause misleading results,
if the full (production and decay)  leading order cross-section 
and the next-to-leading $K$-factor {\it for the 
production process only} are used to 
estimate the full NLO cross-section. 
The $K$-factor ($\mu = m_t$) for the production process
is $323~{\rm fb}/316.9~{\rm fb} \sim 1.02$, so the 
naive 
estimate of the NLO cross-section is 
$1.02 \times \sigma_{\rm LO} \approx 357~{\rm fb}$, 
which is about twenty percent higher than the correct NLO value given in Eq.(\ref{eq26}).  
We emphasize that the ``mixed'' contribution to $t \bar t j$ 
production is a NLO QCD effect, so unless NLO effects 
are properly incorporated into computations of associated production of unstable particles, it is unclear to what 
extent various predictions for cross-sections can be trusted.

In Fig.~\ref{fig3} and \ref{fig4} we show various kinematic distributions for the LHC.
The importance of QCD radiation in decays for various observables 
can be seen from the lower panes. 
We find  that for the LHC, the impact of the QCD radiation 
in the decay is modest; the variable that seems to be most affected is 
$H_\perp$ at small values of the transverse energy. For kinematics distributions  in dilepton invariant mass or in the  relative  azimuthal 
angle of the two leptons, there is a uniform reduction, almost independent 
of $m_{l^+l^-}$ and $\phi_{l^+l^-}$. Finally, we note that 
given the discrepancy between
MC@NLO prediction for the transverse momentum of the $t \bar t$ pair 
and the D0 data \cite{Abazov:2011rq}, 
it is important to measure this distribution 
at the LHC. Thanks to a much higher energy and luminosity, 
the LHC should be able to probe  
a much broader distribution in $p_\perp(t \bar t)$, 
including regions where fixed order QCD computations 
are directly applicable. We show the $p_\perp(t \bar t)$ 
distribution in Fig.~\ref{fig4} and find that
this distribution  receives important  modifications 
 due to radiation in the decay.

\section{Conclusions} 

In this paper, we discussed the computation of NLO QCD corrections to the production 
of a $t \bar t $ pair in association with a hard jet at hadron colliders.
While NLO QCD
corrections to this process
have been considered in the literature several times already, in this article
for the first time, QCD radiative corrections to top quarks decays
are studied, including the possibility that the jet is emitted in the decay stage.
The results reported in this paper 
lead to a complete and fully consistent 
treatment of top quark pair production and decay
in association with a jet at next-to-leading order in perturbative QCD.

While at leading order there is a clear separation into production 
and decay stages, 
at next-to-leading order there appears a new 
contribution where one parton is emitted in the production 
and the other parton in the decay. Since 
this ``mixed'' contribution 
must be supplemented by virtual corrections to ensure infra-red
safety, we find that it can be negative.
This leads to interesting effects
that, to the best of our knowledge, 
have not been discussed in the literature before.
In particular, it is far from clear that a widely used procedure
of {\it estimating} NLO QCD cross-sections by computing leading order cross-sections with  decays and
re-scaling them by $K$-factors obtained from calculations that ignore radiation of jets in the decay
is valid.
In fact, we find that this procedure
accidentally gives an 
accurate estimate of the NLO cross-section for $ t \bar t j$ production 
at the Tevatron but similarly overestimates the NLO QCD
cross-section at the LHC
by twenty percent.   The absence of clear pattern suggests that it is best to 
include QCD radiative corrections to 
 decays of unstable particles into theoretical 
predictions for hard scattering processes.

Jet radiation in the decays can have 
significant impact on kinematic distributions. One such case 
is the $H_\perp$ distribution at the Tevatron  which exhibits significant 
distortion due to radiation in the final state.  While the situation at the LHC 
is less dramatic, even there certain distributions are systematically 
distorted at the ten to twenty percent level. 

We also compare the shape of the transverse momentum distribution
 of a top quark pair recently measured by the D0 collaboration with the 
result of our computation. We combine exclusive $p \bar p \to t \bar t$ 
and inclusive $p \bar p \to t \bar t j$ computations at NLO QCD to describe the 
transverse momentum distribution of $t \bar t$ pair and  find reasonable 
agreement with the results obtained by  D0 collaboration in Ref.~\cite{Abazov:2011rq}.

Recent progress in NLO computations was driven by the idea that perturbative QCD 
can describe hard scattering well, pushing  theorists towards 
providing  realistic descriptions of complicated hard processes 
which can be {\it directly} compared
to experimental data.
Clearly, in the case of heavy short-lived particles such
as top quarks, this implies that NLO QCD computations
should be applied to their decay, including all spin correlations.
All of this can be done in a rather straightforward way 
in the narrow  width approximation which provides a parametric framework
for such studies. 
We have demonstrated how this framework can be used to describe the production 
of $t \bar t$ pairs in association with a jet at hadron colliders. We look
forward to further applying this framework
for the description of both Standard Model and New Physics 
processes at the LHC. 
\\

{\bf Acknowledgments}
This research is supported in part by NSF grants
PHY-0855365 and PHY-0547564, as well as by DOE grant DE-AC02-06CD11357 
and startup funds of Johns Hopkins University.
M.S. is grateful for support from the Director's Fellowship of Argonne National Laboratory.
Calculations reported in this paper were performed on the Homewood
High Performance Cluster of Johns Hopkins University.

\end{document}